\documentclass{webofc}
\usepackage[varg]{txfonts}
\usepackage{hyperref}
\usepackage{url}
\usepackage{placeins}
\hypersetup{colorlinks=true,citecolor=blue,urlcolor=blue,linkcolor=blue}
\graphicspath{{figures/}}

\begin{document}

\title{Kaonic Atom X-ray Spectroscopy: From SIDDHARTA-2 to the EXKALIBUR Program}

\author{%
  \firstname{F.} \lastname{Sgaramella}\inst{1}\fnsep\thanks{\email{francesco.sgaramella@lnf.infn.it}}
  \and \firstname{L.} \lastname{Abbene}\inst{3,1}
  \and \firstname{F.} \lastname{Artibani}\inst{1,4}
  \and \firstname{M.} \lastname{Bazzi}\inst{1}
  \and \firstname{G.} \lastname{Borghi}\inst{5,6}
  \and \firstname{D.} \lastname{Bosnar}\inst{7,1}
  \and \firstname{M.} \lastname{Bragadireanu}\inst{8}
  \and \firstname{A.} \lastname{Buttacavoli}\inst{3,1}
  \and \firstname{A.} \lastname{Clozza}\inst{1}
  \and \firstname{F.} \lastname{Clozza}\inst{1,2}
  \and \firstname{R.} \lastname{Del Grande}\inst{9,1}
  \and \firstname{L.} \lastname{De Paolis}\inst{1}
  \and \firstname{K.} \lastname{Dulski}\inst{1,10,11}
  \and \firstname{C.} \lastname{Fiorini}\inst{5,6}
  \and \firstname{I.} \lastname{Fri\v{s}\v{c}i\'c}\inst{7}
  \and \firstname{C.} \lastname{Guaraldo}\inst{1}\fnsep\thanks{Deceased}
  \and \firstname{T.} \lastname{Hashimoto}\inst{16,17}
  \and \firstname{M. A.} \lastname{Iliescu}\inst{1}
  \and \firstname{M.} \lastname{Iwasaki}\inst{12}
  \and \firstname{A.} \lastname{Khreptak}\inst{10,11,1}
  \and \firstname{S.} \lastname{Manti}\inst{1}
  \and \firstname{J.} \lastname{Marton}\inst{18}
  \and \firstname{P.} \lastname{Moskal}\inst{10,11}
  \and \firstname{H.} \lastname{Ohnishi}\inst{14}
  \and \firstname{F.} \lastname{Principato}\inst{3,1}
  \and \firstname{A.} \lastname{Scordo}\inst{1}
  \and \firstname{M.} \lastname{Silarski}\inst{10}
  \and \firstname{D.} \lastname{Sirghi}\inst{15,1,8}
  \and \firstname{F.} \lastname{Sirghi}\inst{1,8}
  \and \firstname{M.} \lastname{Skurzok}\inst{10,11,1}
  \and \firstname{A.} \lastname{Spallone}\inst{1}
  \and \firstname{K.} \lastname{Toho}\inst{14,1}
  \and \firstname{J.} \lastname{Zmeskal}\inst{13,1}\fnsep\thanks{Deceased}
  \and \firstname{C.} \lastname{Curceanu}\inst{1}%
}

\institute{%
  Laboratori Nazionali di Frascati INFN, Frascati, Italy
  \and Universit\`a degli Studi di Roma Tor Vergata, Dipartimento di Fisica, Roma, Italy
  \and Department of Physics and Chemistry (DiFC)--Emilio Segr\`e, University of Palermo, Palermo, Italy
  \and Universit\`a degli Studi di Roma Tre, Dipartimento di Fisica, Roma, Italy
  \and Politecnico di Milano, Dipartimento di Elettronica, Informazione e Bioingegneria, Milano, Italy
  \and INFN Sezione di Milano, Milano, Italy
  \and Department of Physics, Faculty of Science, University of Zagreb, Zagreb, Croatia
  \and Horia Hulubei National Institute of Physics and Nuclear Engineering (IFIN-HH), M\u{a}gurele, Romania
  \and Faculty of Nuclear Sciences and Physical Engineering, Czech Technical University in Prague, B\v{r}ehov\'a 7, 115 19 Prague, Czech Republic
  \and Faculty of Physics, Astronomy, and Applied Computer Science, Jagiellonian University, Krakow, Poland
  \and Center for Theranostics, Jagiellonian University, Krakow, Poland
  \and RIKEN, Tokyo, Japan
  \and Stefan-Meyer-Institut f\"ur Subatomare Physik, Vienna, Austria
  \and Research Center for Accelerator and Radioisotope Science (RARiS), Tohoku University, Sendai, Japan
  \and Centro Ricerche Enrico Fermi -- Museo Storico della Fisica e Centro Studi e Ricerche ``Enrico Fermi'', Roma, Italy
  \and RIKEN Pioneering Research Institute, Wako, Saitama 351-0198, Japan
  \and RIKEN Nishina Center for Accelerator-Based Science, Wako, Saitama 351-0198, Japan
  \and Technische Universit¨at Wien, Atominstitut, Vienna, Austria
}

\abstract{X-ray spectroscopy of kaonic atoms offers a unique experimental window on the interplay between low-energy strong interaction and precision atomic physics studies. This article reviews the scientific progress achieved with SIDDHARTA-2 at DA$\Phi$NE and discusses how it shapes the next generation measurements. Particular emphasis is given to EXKALIBUR, a multi-facility program designed to perform measurements of kaon--nucleus interactions across the periodic table and to extend kaonic atom spectroscopy toward precision tests of bound-state QED and the charged-kaon mass extraction. The first proposed strong interaction campaign at J-PARC aims to investigate kaonic atoms with solid Li, Be, and B targets using new 1-mm-thick Silicon Drift Detectors and high-resolution TES microcalorimeters, exploiting isotope pairs to isolate nuclear-structure and kaon--multinucleon effects. Measurements of Mg, Al, Si, and S with CdZnTe detectors will broaden this study toward heavier nuclei. EXKALIBUR will establish a systematic experimental framework for testing low-energy strong interaction with strangeness and bound-state QED predictions across a broad range of kaonic atoms.}

\maketitle

\section{Kaonic Atoms as Probes of Strong Interaction and Bound-State QED}
\label{sec:physics-case}

\begin{figure}[htb]
  \centering
  \includegraphics[width=0.60\textwidth,clip]{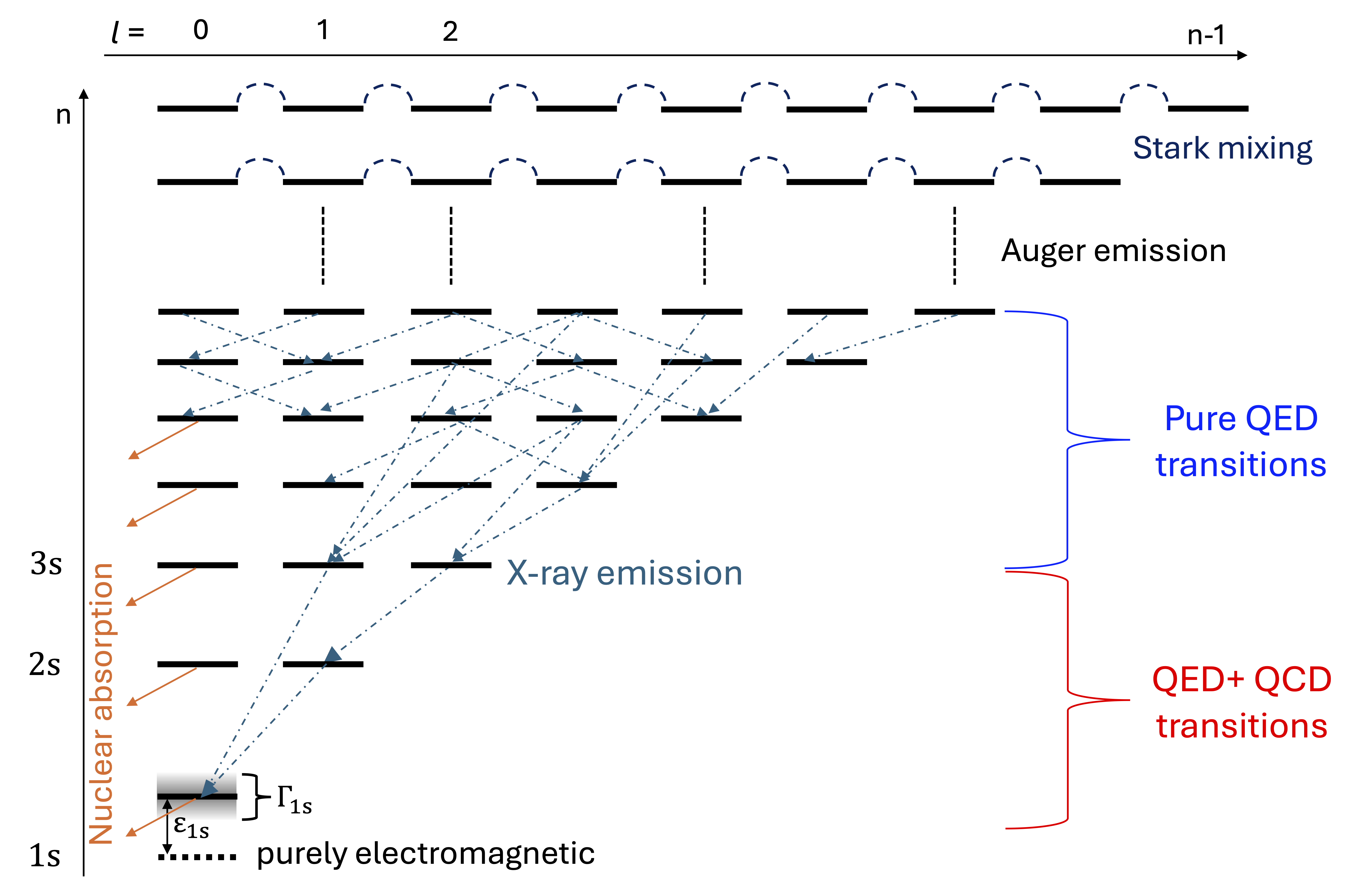}
  \caption{Schematic illustration of the cascade processes in kaonic atoms. The early stages are dominated by Auger emission, followed by radiative X-ray transitions, while nuclear absorption takes place predominantly from the lower energy levels.}
  \label{fig:cascade}
\end{figure}

A kaonic atom is formed when a negatively charged kaon is captured by a nucleus into a highly excited electromagnetic orbit \cite{PhysRev.58.90.2}. The atom subsequently de-excites through the cascade illustrated in Fig.~\ref{fig:cascade}. Auger emission dominates the early stages, when the transition energy slightly exceeds the electronic ionization threshold. For levels closer to the nucleus, X-ray emission dominates, until the kaon is eventually absorbed by the nucleus \cite{Koike:1995sa,Jensen:2004ia,Kalantari:2012isg}. This evolution gives access to two complementary regimes within the same atomic system. Low-lying levels overlap the nuclear wave function and are modified by the strong interaction between the kaon and the nucleus, whereas high-$n$ states remain essentially electromagnetic driven and can be used for precision QED atomic physics studies.

Because the kaon is almost three orders of magnitude heavier than the electron, the corresponding Bohr radius is strongly reduced and the innermost atomic levels approach nuclear dimensions. Their keV-scale binding energies, for the lighter ones, provide access to the kaon-nucleon interaction at threshold, well below the energy range of conventional scattering experiments \cite{Baglin:1973dvg,Rubinstein:1984kf}. Kaonic atoms are, therefore, sensitive probes for investigating the non-perturbative QCD in the strangeness sector \cite{Curceanu:2019uph}.

The strong interaction shifts and broadens the lowest atomic levels with respect to their purely electromagnetic calculated values. The energy shift probes the real part of the kaon–nucleus interaction, while the width reflects its absorptive component. In the lightest systems these observables constrain the threshold $\bar{K}N$ amplitudes and the coupled-channel dynamics associated with the $\Lambda(1405)$ resonance \cite{Lee:1994my,Bernard:2007zu,Kaiser:1995eg}. SIDDHARTA-2 has recently obtained the most precise kaonic hydrogen measurement \cite{SIDDHARTA-2:2026hydrogen} and the first observation of kaonic deuterium X-ray transitions \cite{SIDDHARTA-2:2026deuterium}, providing the experimental basis for extracting the isospin-dependent $K^-N$ scattering lengths.

At higher principal quantum number, the kaonic wave function has negligible overlap with the nucleus and the transitions are governed by electromagnetic effects. The large reduced mass of kaonic atoms results in a much smaller Bohr radius with respect to ordinary atoms, enhancing vacuum-polarization and higher-order QED effects \cite{Paul:2020cnx,Santos:2004bw}. SIDDHARTA-2 has recently measured several transitions in kaonic neon and fluorine providing a proof of concept for precision studies of these effects \cite{Sgaramella:2024klx,SIDDHARTA-2:2025mwc,SIDDHARTA-2:2026mdl}. Kaonic atom spectroscopy thus provides access to both strong interaction effects at low-$n$ and precision electromagnetic effects at high-$n$.

This article reviews the experimental developments that made this combined program possible and focuses on EXKALIBUR, which extends kaonic atom spectroscopy from benchmark systems toward isotope-resolved light nuclei, intermediate-mass targets, BSQED tests, and an improved determination of the charged-kaon mass \cite{Curceanu:2023yuy}.

\section{The Experimental Legacy of SIDDHARTA-2}
\label{sec:siddharta2}

SIDDHARTA-2, installed at the DA$\Phi$NE electron--positron collider at INFN--LNF, established the experimental approach on which the next generation kaonic atoms measurements will be based \cite{Milardi:2009zza,Milardi:2018sih,Sirghi:2023wok,Skurzok:2020phi}. Low-momentum kaon pairs from $\phi$-meson decays are identified by a coincidence trigger with a time resolution of about 350~ps FWHM. The negative kaons are then slowed by a degrader and stopped in a cryogenic gaseous target . A central technological breakthrough was the development of a large-area, segmented Silicon Drift Detector (SDD) system, capable of providing spectroscopic and timing information simultaneously. The system comprises 384 independent channels surrounding the target, with an energy resolution of approximately 150~eV FWHM at 6.4~keV and a timing resolution of about 450~ns FWHM \cite{Miliucci:2021wbj,Miliucci:2022lvn}. This combination enables precise X-ray spectroscopy while suppressing asynchronous background through time correlations with the kaon trigger, providing a key capability for measurements of weak kaonic atom transitions.\\
Dedicated veto systems were also developed for kaonic atom spectroscopy, combining active charged-particle rejection with fiducial-volume selection to suppress backgrounds associated with kaon stops outside the gaseous target \cite{Sirghi:2023wok,Tuchler:2023ozy,Bazzi:2013kwa}. The capabilities of this apparatus were demonstrated by the first observation of kaonic-deuterium X-ray transitions, one of the most demanding measurements in kaonic atom spectroscopy because of the extremely small radiative yield and the large intrinsic width of the lines \cite{SIDDHARTA-2:2026deuterium}.

These developments provide the technological foundation for future kaonic atom experiments. Scalable segmented SDD arrays, optimized stopping distributions and layered background suppression can be adapted to different targets and beam conditions. In particular, the SIDDHARTA-2 SDD architecture has evolved toward 1-mm-thick monolithic arrays that preserve segmentation and timing capability while increasing the detection efficiency at the higher X-ray energies \cite{Toscano:2024vwf}. Together with CdZnTe detectors for harder X-rays and TES microcalorimeters for transitions requiring substantially higher energy resolution, the SIDDHARTA-2 experiment provides the basis for the EXKALIBUR program.

\begin{figure}[htb]
  \centering
  \includegraphics[width=0.60\textwidth,clip]{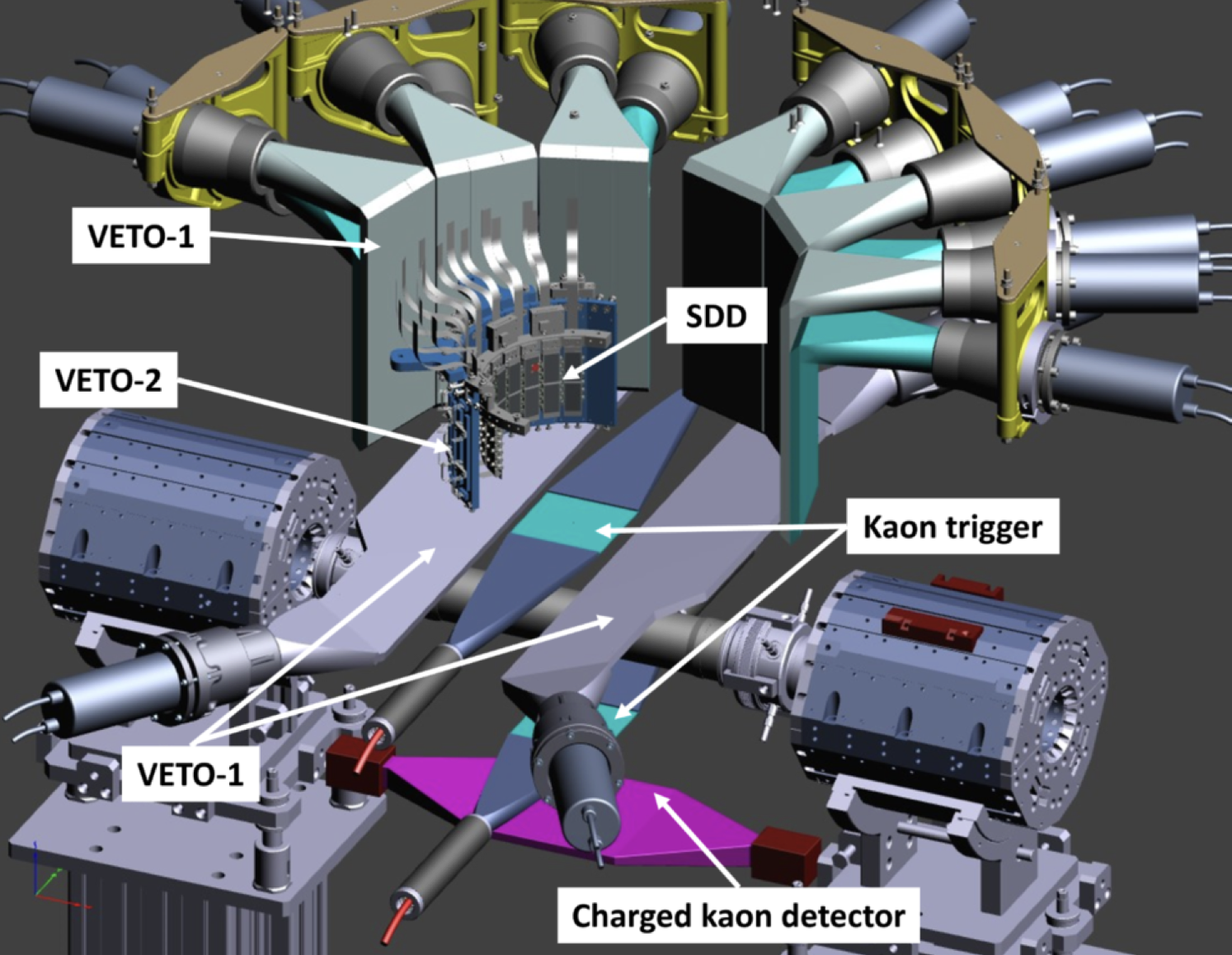}
  \caption{Schematic view of the SIDDHARTA-2 apparatus, showing the kaon trigger, the Silicon Drift Detectors (SDDs), and the veto detectors for background suppression and event selection. Figure adapted from Ref.~\cite{Sirghi:2023wok}.}
  \label{fig:setup}
\end{figure}

\section{EXKALIBUR: a Modular Program for Future Kaonic Atoms Spectroscopy}
\label{sec:exkalibur}

EXKALIBUR (EXtensive Kaonic Atoms research: from LIthium and Beryllium to URanium) builds on the detector and analysis experience accumulated with SIDDHARTA-2 and extends it into a systematic program covering selected targets across the periodic table \cite{Curceanu:2023yuy,Manti:2025exkalibur}. Unlike SIDDHARTA-2, which was developed specifically for the DA$\Phi$NE interaction region, EXKALIBUR is designed for a broader experimental program at both DA$\Phi$NE and J-PARC. The apparatuses are adaptable to the beam and background conditions of each facility, rather than being a single setup transferred unchanged from one facility to the other. They, nevertheless, share important detector developments, in particular 1-mm-thick SDD arrays and CdZnTe spectrometers, whereas TES microcalorimeters are reserved for measurements at J-PARC. The choice of detector technology is guided by the X-ray energy range and the energy resolution required to address each scientific objective.

The scientific program is structured around three interconnected research questions: how does the low-energy $K^-$--nucleus interaction evolve from one-nucleon and two-nucleon systems to nuclei in which multi-nucleon absorption and density-dependent optical potentials become essential?; can high-$n$ transitions provide competitive tests of BSQED in strong electromagnetic fields while controlling nuclear-size and electron-screening corrections?; can a dedicated measurement resolve the long-standing charged-kaon mass puzzle?

\subsection{The First Phase: a Solid Li Be B Target Surrounded by 1-mm SDDs}
\label{sec:exkalibur-first-phase}

\subsubsection{The setup}
\begin{figure}[htb]
  \centering
  \includegraphics[width=0.5\textwidth,clip]{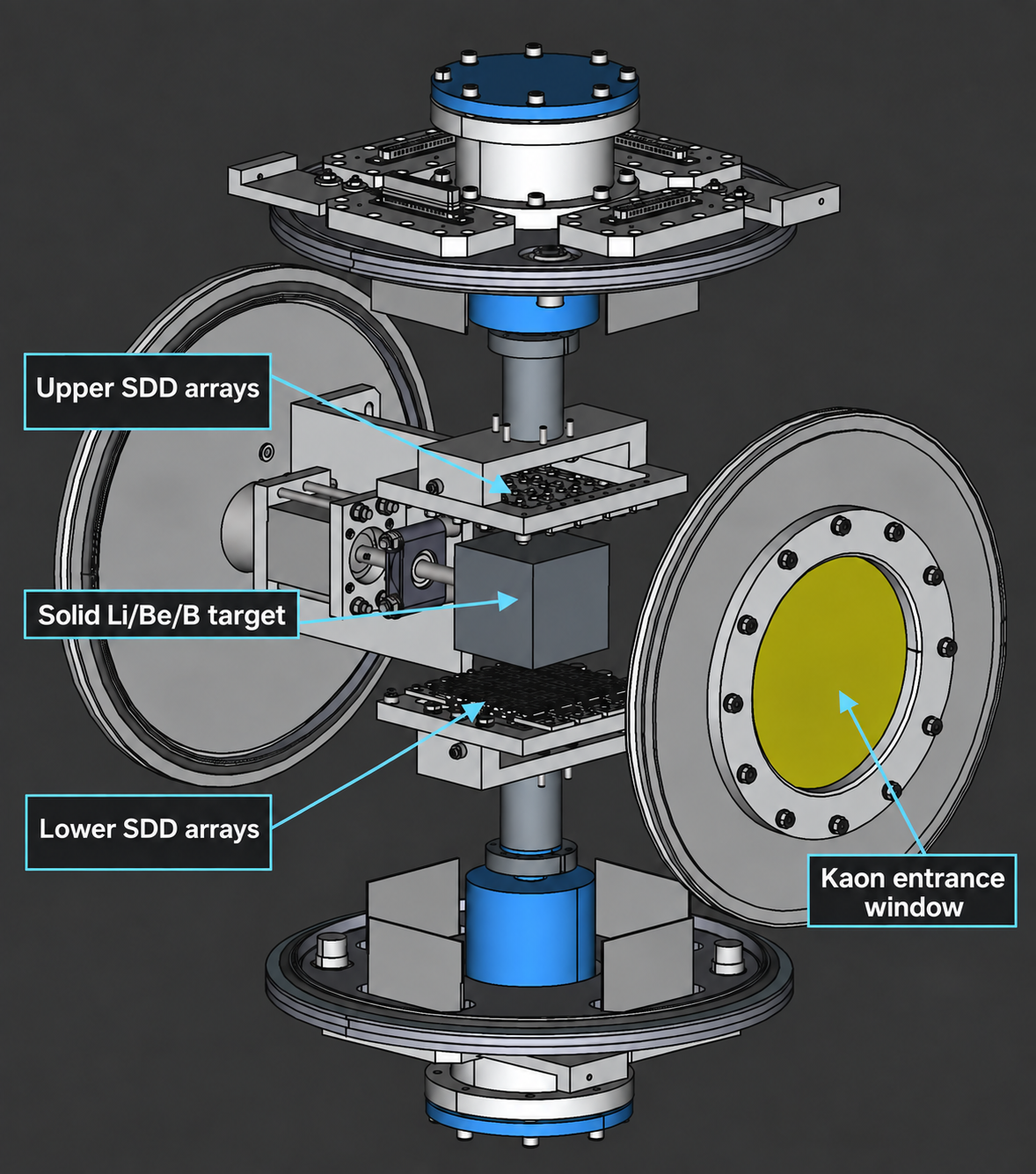}
  \caption{Conceptual layout of the first EXKALIBUR phase at J-PARC for solid Li, Be, and B targets. Kaons enter through the lateral circular window and stop in the central solid target. Eight new-generation 1-mm-thick SDD matrices are installed above the target and eight below it, for a total of 16 matrices, 128 readout channels, and 81.92~cm$^2$ of active area.}
  \label{fig:exkalibur-setup}
\end{figure}

The first EXKALIBUR phase is a dedicated J-PARC experiment based on a compact setup with interchangeable solid Li, Be, and B targets and 1-mm SDDs. The schematic view shown in Fig.~\ref{fig:exkalibur-setup} applies the experimental methods established by SIDDHARTA-2 based on SDDs, kaon tagging, detector timing, vacuum operation, calibration, and background rejection, to the J-PARC beam environment, with a geometry optimized for light solid materials.
\FloatBarrier

The detector thickness is a decisive upgrade. Increasing the SDD active thickness from 450~$\mu$m to 1~mm improves the detection efficiency at the higher X-ray energies relevant to the Li Be B program. The new monolithic arrays retain the $8\times8$~mm$^2$ pixel geometry and eight-pixel matrix format \cite{Toscano:2024vwf}, providing approximately 90\% efficiency at 15~keV and 20\% at 30~keV. This extends the accessible energy range toward the 3d--2p transitions of Li, Be, and B.

\subsubsection{Light Kaonic Atoms and the Kaon--Multinucleon Interaction}
\label{sec:exkalibur-light}

The primary strong interaction targets are $K^-{}^6$Li, $K^-{}^7$Li, $K^-{}^9$Be, $K^-{}^{10}$B, and $K^-{}^{11}$B. Their low-lying atomic orbits overlap the nucleus, where the one-nucleon $\bar{K}N$ amplitude alone is insufficient to describe the kaon-nucleus interaction. Modern optical-potential analyses require density-dependent multinucleon terms, while microscopic calculations seek to derive $K^-NN(N)$ absorption from the in-medium dynamics \cite{Friedman:1994hx,Friedman:2016rfd,Obertova:2022wpw}. Precise line shifts, widths, and yields in Li, Be, and B, therefore, test the transition from few-body dynamics to a genuinely density-dependent kaon--nucleus interaction.

The isotope pairs are especially valuable. Comparing $^6$Li with $^7$Li and $^{10}$B with $^{11}$B changes the neutron number while keeping the charge and much of the atomic structure fixed. QED contributions, and several cascade uncertainties are then correlated and partly cancel in the isotope difference. The remaining variation is sensitive to the isovector part of the optical potential, to nuclear-surface structure, and to clustering effects. 

Table~\ref{tab:exkalibur-strong-goals} lists the calculated electromagnetic transition energies and tthe target precisions for the shift and width for the first phase of the EXKALIBUR program. The predicted strong interaction effect grows rapidly from lithium to boron. The precision goals therefore scale with the intrinsic line broadening. An uncertainty of 1~eV on the lithium shift and 2~eV on its width would make the lithium isotope pair a particularly stringent test of the strong interaction effect in light systems. For beryllium the corresponding goals are 3 and 5~eV, while for the much broader boron line they are 8 and 20~eV.
\begin{table}[htb]
  \centering
  \caption{Calculated transition energies and expected strong interaction observables for selected EXKALIBUR targets. All energies, shifts, widths, and uncertainties are expressed in eV.}
  \label{tab:exkalibur-strong-goals}

  \small
  \renewcommand{\arraystretch}{1.15}
  \setlength{\tabcolsep}{5pt}

  \begin{tabular}{ccc|cc|cc}
    \hline
    \textbf{Atom} &
    \textbf{Transition} &
    \textbf{Energy} &
    \multicolumn{2}{c|}{\textbf{Shift}} &
    \multicolumn{2}{c}{\textbf{Width}} \\
    & & &
    \textbf{Prediction} & \textbf{Target precision} &
    \textbf{Prediction} & \textbf{Target precision} \\
    \hline

    $K^-{}^{6}$Li
      & 3d--2p & 15152
      & $-3$   & 1
      & 44     & 2 \\

    $K^-{}^{7}$Li
      & 3d--2p & 15329
      & $-3$   & 1
      & 54     & 2 \\

    $K^-{}^{9}$Be
      & 3d--2p & 27708
      & $-37$  & 3
      & 247    & 5 \\

    $K^-{}^{10}$B
      & 3d--2p & 43567
      & $-170$ & 8
      & 718    & 20 \\

    $K^-{}^{11}$B
      & 3d--2p & 43767
      & $-179$ & 8
      & 757    & 20 \\

    \hline

    $K^-{}^{24}$Mg
      & 4f--3d & 90553
      & -27 & 10
      & 214 & 20 \\

    $K^-{}^{27}$Al
      & 4f--3d & 106567
      & -76 & 10
      & 442 & 20 \\

    $K^-{}^{28}$Si
      & 4f--3d & 123719
      & -130 & 10
      & 800 & 20 \\

    $K^-{}^{32}$S
      & 4f--3d & 162071
      & -500 &20
      & 2000 & 50 \\
    \hline
  \end{tabular}
\end{table}

Intrinsic widths of only a few tens of eV, however, are difficult to separate from the SDD response. For this reason, selected lines will also be studied with TES microcalorimeters in a second phase. Their much narrower response can distinguish the $^6$Li/$^7$Li doublet directly and measure the intrinsic width. In this way, SDD and TES measurements are complementary: the SDD establish a high-statistics survey and the TES target the most resolution-sensitive observables.

\subsection{Intermediate-Mass Systems with CdZnTe Detectors}
\label{sec:exkalibur-czt}

The kaon–nucleus interaction will be investigated over a wider nuclear-mass range through measurements of Mg, Al, Si, and S. The Mg, Al, and Si targets will probe the evolution of the density-dependent optical potential with increasing nuclear mass. Sulfur provides a particularly important benchmark, as previous measurements of its low-lying transitions are mutually inconsistent and are not simultaneously reproduced by optical-potential models \cite{Friedman:1994hx}. A new measurement of the transition energies and widths could clarify whether this discrepancy originates from experimental systematics or from an incomplete description of the kaon–multinucleon interaction. 

The corresponding 4f--3d transition energies, listed in the lower part of Table~\ref{tab:exkalibur-strong-goals}, span approximately 91--162~keV. Silicon detectors become inefficient in this domain, whereas the high atomic numbers and density of CdZnTe provide excellent detection efficiency and good energy resolution at room temperature \cite{Scordo:2023per, Abbene:2023vhm, Abbene:2023ogi}. Quasi-hemispherical CdZnTe prototypes developed for kaonic atom spectroscopy reach nearly 100\% efficiency around 63~keV and about 70\% around 162~keV, with compact front-end electronics and no cryogenic system \cite{Abbene:2023ogi,Abbene:2023vhm}. Prototype modules have already been tested in the DA$\Phi$NE environment \cite{Scordo:2023per} in exploratory measurements of Al, Cu, and F \cite{Artibani:2026wvc}, and validating this technology in the view of dedicated measurements with Mg, Al, Si, and S targets at J-PARC.

\subsection{BSQED and the Charged-Kaon Mass with SDDs and TES Microcalorimeters}
\label{sec:exkalibur-bsqed}

The EXKALIBUR program includes the study of high-$n$ circular transitions, which are predominantly governed by electromagnetic effects. Because their wave functions remain far from the nucleus, the strong interaction shift is negligible, while vacuum polarization and other BSQED contributions remain enhanced due to the large reduced mass of the kaonic atom system. Compared to highly charged ions, a suitable kaonic atom can reduce nuclear-size and polarization uncertainties and, with respect to muonic atoms, the larger reduced mass enhances QED sensitivity. Moreover, because the kaon has spin zero, there is no kaon-spin fine or hyperfine structure to disentangle \cite{Paul:2020cnx,Santos:2004bw}. A sequence of transitions across several $Z$ values can therefore separate the scaling of first- and second-order QED terms from recoil, screening, and finite-size corrections.

The same transitions encode the charged-kaon mass through the reduced mass of the atom. The present value, $m_{K^\pm}=493.677\pm0.013$~MeV/$c^2$, is affected by a disagreement of approximately 60 keV between the two historically most precise measurements \cite{Gall:1988ei,Denisov:1991pu,ParticleDataGroup:2024cfk}. 
The sub-eV statistical precision achieved by SIDDHARTA-2 for selected kaonic neon transitions identifies neon as a particularly promising system for addressing the existing charged-kaon mass discrepancy \cite{Sgaramella:2024klx,Manti:2025exkalibur}. For this reason, EXKALIBUR  envisages a dedicated kaonic neon measurement with an apparatus combining SDD and TES detectors, optimized for high efficiency and precise energy calibration. Keeping the energy-scale uncertainty below 0.1~eV and the total transition-energy uncertainty below about 0.2~eV would translate into a charged-kaon mass uncertainty below 10~keV \cite{Manti:2025exkalibur,SIDDHARTA-2:2025mwc}.
SDDs and TES microcalorimeters provide complementary capabilities for this measurement. Thanks to their large active area and timing performance, SDDs can accumulate useful statistics rapidly, allowing the overall apparatus performance, line intensities, satellite components, cascade effects, and background to be characterized on short timescales. TES arrays have an active area roughly two orders of magnitude smaller, but offer an energy resolution improved by a comparable factor, making them particularly suited to resolving closely spaced spectral structures. This becomes crucial for the extraction of the charged-kaon mass, where unresolved satellite lines or imperfectly modelled screening corrections can mimic an energy shift at the required precision. A combined SDD--TES analysis therefore exploits the high-statistics spectral characterization provided by the SDDs together with the superior energy resolution of the TES detectors. The 50- and 100-keV TES designs considered for the J-PARC program are summarized in Table~\ref{tab:tes-options}; their operation has been demonstrated in accelerator environments, and the array configuration can be optimized according to the transition energy \cite{HEATES:2016giw,Hashimoto:2016jnk,J-PARCE62:2020klj}.

\begin{table}[htb]
  \centering
  \caption{Representative TES configurations considered for the EXKALIBUR precision program \cite{E57Status:2026}. Areas refer to the full array.}
  \label{tab:tes-options}
  \small
  \setlength{\tabcolsep}{3.5pt}
  \begin{tabular}{lccc}
    \hline
    \textbf{Parameter} & \textbf{10-keV TES} & \textbf{50-keV TES} & \textbf{100-keV TES} \\
    \hline
    Absorber & Bi & Au + Bi & Sn \\
    Pixels & 240 & 96 & 96 \\
    Total area [mm$^2$] & 21.2 & 43.1 & 162 \\
    Resolution (FWHM) & 5~eV at 6~keV & 20~eV at 45~keV & 80~eV at 100~keV \\
    Representative efficiency & 86\% at 6~keV & 90\% at 15~keV & 98\% at 50~keV \\
    \hline
  \end{tabular}
\end{table}
\FloatBarrier

\section{Outlook and Conclusions}
\label{sec:summary}

The recent SIDDHARTA-2 results mark an important step toward a broader use of kaonic-atom X-ray spectroscopy as a quantitative probe of both low-energy strong interaction and precision bound-state QED. The measurements of kaonic hydrogen and deuterium constrain the elementary $\bar{K}N$ interaction at threshold, while the kaonic neon results demonstrate the precision achievable for electromagnetically dominated transitions. Together with the development of large-area timed SDD arrays, these results provide the experimental basis for extending kaonic atom spectroscopy beyond the systems studied so far.

EXKALIBUR is designed to systematically carry out this extension. Measurements of Li, Be, and B at J-PARC will determine strong interaction shifts and widths in light nuclei and exploit isotope pairs to investigate nuclear-structure, isospin, and kaon--multinucleon effects. The extension to Mg, Al, Si, and S with CdZnTe detectors will test the evolution of the kaon--nucleus optical potential toward larger nuclear masses and revisit regions where existing measurements remain incomplete or inconsistent. High-resolution TES measurements will complement the larger-area detectors whenever intrinsic widths, isotope separations, or nearby spectral components require substantially better energy resolution.

At the same time, high-$n$ transitions provide access to a complementary regime in which strong interaction effects are suppressed and electromagnetic contributions dominate. Their measurement can therefore be used to test BSQED calculations and to improve the determination of the charged-kaon mass, with SDD and TES detectors providing complementary control of statistics, spectral structure, and systematic uncertainties. EXKALIBUR thus opens the possibility of studying, within the same experimental program, both the evolution of the kaon--nucleus interaction across the periodic table and the precision spectroscopy of electromagnetically dominated kaonic atom transitions.

\section*{Acknowledgement}
We thank C. Capoccia from LNF-INFN and H. Schneider, L. Stohwasser, and D. Pristauz-Telsnigg from Stefan Meyer-Institut for their fundamental contribution in designing and building the SIDDHARTA-2 setup. We also thank the DA$\Phi$NE staff for the excellent working conditions and permanent support. Special thanks to Catia Milardi for her continued support and contribution during the data taking. Part of this work was supported by the INFN (KAONNIS project); the Austrian Science Fund (FWF): [P24756-N20 and P33037-N]; the Croatian Science Foundation under the project IP-2022-10-3878; the EU STRONG-2020 project (Grant Agreement No. 824093); the EU Horizon 2020 project under the MSCA (Grant Agreement 754496); the Japan Society for the Promotion of Science JSPS KAKENHI Grant No. JP18H05402; the Polish Ministry of Science and Higher Education grant No. 7150/E-338/M/2018 and the Polish National Agency for Academic Exchange (grant no PPN/BIT/2021/1/00037); the EU Horizon 2020 research and innovation programme under project OPSVIO (Grant Agreement No. 101038099). This article/publication is based upon work from COST Action CA24131-ENRICH, supported by COST (European Cooperation in Science and Technology, http://www.cost.eu/).

\bibliography{ref}

\end{document}